# Patterns of Nanoflare Storm Heating Exhibited by an Active Region Observed with SDO/AIA

Nicholeen M. Viall and James A. Klimchuk

NASA Goddard Space Flight Center, Greenbelt, MD

**Abstract**

It is largely agreed that many coronal loops---those observed at a temperature of about 1 MK---are bundles of unresolved strands that are heated by storms of impulsive nanoflares. The nature of coronal heating in hotter loops and in the very important but largely ignored diffuse component of active regions is much less clear. Are these regions also heated impulsively, or is the heating quasi steady?  The spectacular new data from the Atmospheric Imaging Assembly (AIA) telescopes on the Solar Dynamics Observatory (SDO) offer an excellent opportunity to address this question. We analyze the light curves of coronal loops and the diffuse corona in 6 different AIA channels and compare them with the predicted light curves from theoretical models. Light curves in the different AIA channels reach their peak intensities with predictable orderings as a function the nanoflare storm properties. We show that while some sets of light curves exhibit clear evidence of cooling after nanoflare storms, other cases are less straightforward to interpret.  Complications arise because of line-of-sight integration through many different structures, the broadband nature of the AIA channels, and because physical properties can change substantially depending on the magnitude of the energy release. Nevertheless, the light curves exhibit predictable and understandable patterns consistent with impulsive nanoflare heating.

## 1. Introduction

Understanding the mechanisms through which the solar corona is heated to temperatures of greater than one million Kelvin remains as one of the prominent challenges of solar physics. In order to make progress on this very difficult problem, we focus on the heating of a quiescent active region and use the spectacular new observations made with the Atmospheric Imaging Assembly (AIA) telescopes (Boerner et al. 2011; Lemen et al. 2011) on the Solar Dynamics Observatory (SDO). In Figure 1 we show this active region, NOAA AR 11082, observed on 2010 June 19, in the northern hemisphere, near central meridian. We display the same (~240"x240") region in 6 EUV channels all on a linear scale, observed at 8.4 UT. From left to right and top to bottom, we display channel 131 Å, 171 Å, 193 Å, 211 Å, 335 Å and 94 Å. The channels corresponding to these images are nominally in temperature order, and we list the temperature of peak sensitivity in the upper left corner of each image.

Historically, active regions have been described in the literature in terms of five main components: the core region, which contains hot (> 2 MK) plasma (Warren et al. 2010); 'extended' warm (1 MK) coronal loops which surround the core (Klimchuk 2006; Warren et al. 2010); the diffuse emission, which typically refers to background emission where there are no distinct 'loops' present (Klimchuk 2006); moss emission from the foot points of hot core loops (Martens et al. 2000); and warm fan structures at the active region periphery (Schrijver et al. 1999; Ugarte-Urra et al. 2009). We note that these terms are not always used in the literature, and sometimes they are used to mean different things, as is common when nomenclature is introduced into a field. In particular, sometimes the terms 'AR core' and 'diffuse emission' are used interchangeably, as are the terms 'quiet sun' and 'diffuse emission'. This is in contrast to our usage of 'diffuse emission', which is any EUV emission within the active region that is not associated with distinguishable loops or loop foot points. For the purposes of the analysis we present here, we focus on the first three components, and explicitly avoid the last two components. These active region components are all visible in Figure 1. Further, because the thermal distribution of the plasma is not uniform, the active region has a different appearance in these 6 different AIA channels. For example, the 171 channel is most characterized by 'extended' warm loops, while the 335 channel is most characterized by core emission. In order to understand the dominant heating mechanism of the active region as a whole, it is important to know if each of these components is heated in a fundamentally different way, or if the entire active region is accurately described by heating through one mechanism (e.g., impulsive nanoflares).

Though at first glance the images in Figure 1 are dominated by different features, note that the components of the active region are not mutually exclusive. While 'extended' warm loops are most prevalent in the volume outside of the core, there are warm loops found within the core as well. Similarly, diffuse emission is present everywhere. We show in Section 3 that the diffuse component actually dominates at all locations in both warm and hot channels. Given the increased understanding of active regions recently, along with the research we present in this paper, we will argue in the conclusion of this paper for a different vocabulary which more accurately describes the salient properties of active regions. Namely, we suggest that 'loops' describes any distinguishable, discrete intensity enhancement whether it occurs in the 'extended' regions or in the core. Likewise, we suggest that emission lacking a discrete intensity enhancement is 'diffuse' whether it occurs in the core or surrounding regions, and whether it occurs in hot or warm temperatures.

For this paper, we define a coronal loop as an observational feature, likely composed of tens to hundreds of coronal strands. A coronal strand, in contrast to a coronal loop, is a physical feature, rather than an observational one. It is a mini flux tube for which the heating and plasma properties are roughly constant over the cross section; the cross sectional area is likely well below the instrument resolution. We define a nanoflare to be an impulsive release of energy on a coronal strand. Though the concept of nanoflare heating often refers to the theory put forth by Parker (1983, 1988), in our usage here we are not implying a specific mechanism. Lastly, a nanoflare storm is a collection of nanoflares occurring in nearby, but physically separate, strands over a finite window in time.

The idea that loops are heated by nanoflares has been popular for some time (Klimchuk 2006 and references cited therein), and it is now largely agreed upon that warm loops are explained by nanoflare storms with durations of several hundred to several thousand seconds (e.g. Warren et al. 2002; Warren et al 2003; Winebarger et al. 2003; Winebarger and Warren 2005; Klimchuk 2009), although see Mok et al. (2008) for an alternative explanation. In this scenario, each of the many hundreds to thousands of strands in the loop is heated impulsively and the heating recurrence time on a single strand is longer than a cooling time. Such loops are then expected to be seen sequentially in cooler channels with time, a pattern confirmed in X-ray and EUV observations (Warren et al. 2002; Winebarger et al. 2003; Winebarger and Warren 2005; Ugarte-Urra et al. 2006; Ugarte-Urra et al. 2009). On the other hand, in the so-called hot cores of active regions, some believe the heating to be predominantly steady, meaning that the heating recurrence time on a strand is much less than the cooling time, and that the emission is exclusively hot. In fact, a dearth of warm emission in active region cores is one argument made in support of steadily heated cores (Warren et al. 2011; Winebarger et al. 2011; although see Tripathi et al. 2011). If the entirety of the hot core is heated in a quasi-steady manner, then all of the plasma should be hot, as each coronal strand is reheated before it has a chance to cool.

In this paper we examine the heating mechanism in the core of this active region. First, we examine the validity of the view that active region cores are exclusively hot. Second, we look at exactly how much total emission is in the form of discrete loops, and how much is in the form of diffuse emission. Lastly, we investigate the heating mechanism of this active region core through the analysis of three locations. Two are locations where loops are clearly identifiable in the 171 image, and one is a diffuse region in the AR core. For each of the three locations, we compare light curves in all 6 channels with a model of nanoflare storm heating, and show that all of these locations are consistent with impulsive nanoflare storm heating, and inconsistent with quasi-steady heating.

**2. Data and Modeling**

We use level 1.5 SDO/AIA data for 2010 June 19 obtained through the AIA cutout service. We use 12 hours of full image resolution (0.6 arc seconds/pixel) data for the 131, 171, 193, 211, 335 and 094 channels at a 70 second cadence. Level 1.5 data have been flat fielded, despiked, and all channels are co-aligned and co-scaled. In Figure 2, we show the response functions of these 6 different channels, all normalized to their own maxima. In our discussion of Figure 1, we indicate the temperature to which each channel is most sensitive; however, all of the channels are in fact sensitive to a range of temperatures. Some of the channels are double peaked with substantial sensitivities at both very high and low temperatures. It is crucial that this feature of the filters is taken into account when analyzing light curves observed in these filters. Furthermore, it is important to keep in mind that these response functions have been calculated assuming that the emitting plasma is in ionization equilibrium. This will not be the case for the extremely hot and tenuous plasma that is present shortly after a nanoflare occurs (Bradshaw and Cargill 2006; Reale and Orlando 2008). Since the emission from highly ionized species is diminished in this situation, the high temperature peaks of the response functions of the 131 channel and, to a lesser extent, the 94 channel, predict stronger emission than can actually be expected (Bradshaw and Klimchuk 2011).

In the analysis we present in this paper, we compare all of the observed light curves with light curves predicted using the Enthalpy-Based Thermal Evolution of Loops (EBTEL) model (Klimchuk,

Patsourakos, and Cargill 2008). EBTEL is a 0-D hydrodynamic model that describes the evolution of average temperature, pressure and density along a coronal strand as a function of time. EBTEL computes the differential emission measure as a function of time that results for each strand. We use EBTEL to model multi-stranded loops that are heated by nanoflare storms. EBTEL predicts various observational signatures as the nanoflare storm properties change. For example, nanoflare storm duration is positively correlated with the observed lifetime and thermal distribution of a coronal loop (Klimchuk 2009). Though individual strands cool in a typical cooling time, new strands will be heated for as long as the nanoflare storm persists and contribute to the overall intensity of each channel as they cool. As the storm duration increases, more strands will be at different phases of heating and subsequent cooling, therefore the width of the resulting thermal distribution also increases with increasing storm duration. A loop produced by a long duration storm will likely be observable in more channels at a single instance in time.

Another property predicted to change as the details of the nanoflare storm change is the intensity ratio in a pair of channels, which is a function of the magnitude of a nanoflare storm. There is more emission at hotter temperatures in a stronger storm, and more emission in warm temperatures in less intense storms (Klimchuk et al. 2008). This property of nanoflare storms, when coupled with broad and/or double peaked response functions of the AIA channels, leads to very different light curves as a function of storm magnitude. To predict specifically what each AIA channel will observe for a given set of nanoflare storm parameters, we fold the total storm emission predicted by EBTEL through the AIA instrument response functions shown in Figure 2. In Figure 3, we show two nanoflare storms of the same 500s duration coupled with the AIA response functions. The individual nanoflares have a triangular rise and fall lasting 500 s, and they are initiated uniformly over a 500 s storm window. The strands have a full length of $1.5 \times 10^{10}$ cm and an initial electron density of $3.2 \times 10^7$ cm$^{-3}$. We show results for a weak storm in Figure 3a, while the one in 3b is strong, having 10 times as much energy per nanoflare. In the case of the weak storm, the cooler components of the 94 and 131 channels dominate the light curve, while in the case of the strong storm, the hotter component dominates and 131 and 94, causing them both to peak before the other channels. However, we once again caution that the predicted hot component may be artificially strong due to ionization nonequilibrium effects. Given these two examples, it seems the light curves due to nanoflare storm heating peak in 335 first followed by 211, 193 and then 171, while 94 and 131 can reach peak intensity at different relative timings, depending on the storm magnitude.

**3. Results**

To begin with, we test the view that active region cores are exclusively hot. In Figure 4 we plot an intensity slice diagonally through the center of the active region core in channels 335 (a hot channel) and 171 (a warm channel). This slice is explicitly chosen to run through the center of the AR core, roughly parallel to the magnetic neutral line, while avoiding the moss, and corresponds to the diagonal line drawn on the 335 and 171 images shown in Figure 1. We express the intensities as emission measures, EM ($cm^{-5}$), under the assumption that the plasma is isothermal at the temperature of peak channel sensitivity (0.8 MK for 171 and 2.5 MK for 335). The 171 values are multiplied by 52 to be on the same scale as the 335 values.

We indicate in Figure 4 the locations in the active region which are traditionally associated with the 'extended' warm loop emission (pixels 0-270; 320-400) versus that of the hot core (pixels 270-320). The 335 profile exhibits clear differences between the core and the extended loops region, namely there is much higher emission in the core. Likewise, the core emission exceeds all emission in the extended loops region in the 171 slice, though the intensity fall-off is less steep than that of 335. In the 171 slice there is plenty of warm emission in the 'hot' core, both in the form of discrete enhancements (loops) as well as in diffuse emission. It is not the case that hot and warm emission are mutually exclusive.

There is always an ambiguity when performing temperature diagnostics with an instrument such as AIA that does not isolate individual spectral lines. We cannot be certain that the 171 emission comes primarily from warm plasma or that the 335 emission comes primarily from hot plasma. It is significant, however, that this assumption leads to a ratio of hot-to-warm emission measures in the core (approximately 50) that agrees well with what Tripathi et al. (2011) found in the cores of two active regions observed by the EUV Imaging Spectrometer (EIS) on Hinode. We note that Warren et al (2011) and Winebarger et al. (2011) found significantly larger ratios in two other active regions also observed by EIS.

It is possible that some emission in the 171 slice which we have attributed to the 'core' is in fact due to extended loops which actually overlie the core along the same line of sight. A simple comparison of the total EM of 171 in the core to the extended loop region along this slice suggests that extended loops located in the foreground of the core can make up no more than 50% of the total core emission. Of course, the identified extended loop regions along this slice are not the same extended loops which are in the foreground of the core. To estimate the emission coming from these latter loops, we examine the two regions in Figure 5 marked with white rectangles, which we believe to represent the legs of loops that directly overly the core. We follow the method of Tripathi et al. (2011) to estimate the warm foreground contribution, taking into account the effects of gravitational stratification, which will tend to reduce loop top emission relative to emission from the loop legs. The mean emission measure (EM) for the lower right region is $5.54 \times 10^{26}$ cm$^{-5}$, and the mean EM for the upper left region is $7.14 \times 10^{26}$ cm$^{-5}$. We estimate that the footpoints of these AR loops are separated by 76", or about $5.7 \times 10^4$ km. Loop top emission from loops of this size are a factor of 0.14-0.4 reduced relative to the loop legs, depending on whether the plasma is hydrostatic, or super hydrostatic. Per this calculation, the total foreground EM is between $0.78$-$2.9 \times 10^{26}$ cm$^{-5}$, where the range is due to the range in the reduction factor and the different mean EM of the foreground regions. Therefore, warm foreground plasma is likely contributing between 15-57% of the mean EM shown in the core slice in 171, consistent with our conclusion that there is a large quantity of warm emission in the core relative to the more traditionally thought of warm emission region (the 'extended' loop region).

A second observation visible in Figure 4 is that in all regions, in both channels, loop emission is not the dominant emission. The diffuse emission comprises the majority of the emission in the entire active region, in all channels. Even the brightest loops are at most ~35% above the diffuse emission, if the diffuse component around the loop is taken to be the mean of some neighboring pixels. There are many loops that are clearly identifiable and distinguishable by eye in an image, which turn out to be a mere 10-20% of the total emission when we quantitatively examine the intensity profiles. In Figure 4 we highlight two loops visible in both 335 and 171 to illustrate this point. The loop located near ~pixel 297 is extremely bright, perhaps the brightest in either image, yet it is still only ~35% of the total emission in 335 and only ~30% in channel 171. Near pixel 309 is another loop, also a distinct enhancement over the diffuse background in both channels, but only 10% of the total emission. DeForest et al. (2009) show that deconvolving TRACE 171 images can enhance the contrast of bright features somewhat. In their examples, loops which are a 30% increase over the diffuse emission in the original image are as great as 35-40% increase over the diffuse emission after deconvoling the image. Importantly, in either estimation, the diffuse component surrounding AR loops comprises a majority of the total emission in the AR.

Now that we have established that there is warm and hot emission collocated in this active region core, and furthermore that the diffuse component is a very significant contributor to the total emission of this active region, we investigate the temporal behavior of three locations in the active region core. Two locations are loops in the core, and one is of a diffuse emission in the core, lacking distinguishable loops in any channel. We test whether the light curves exhibit behavior consistent with steady heating, or whether their behavior is consistent with the predictions of the EBTEL model for impulsive nanoflare storm heating.

3.1 Location 1: Loop Emission

In Figure 6 we display the core region of the 171 and 335 images (as shown in Figure 1, on a linear scale) magnified by a factor of five. We identify a loop visible in both 171 and 335, indicated with arrows. The 171 image was taken at 8.4 UT, the same as the data used in Figures 1, 4 and 5, while the 335 image was taken earlier at 8.2 UT, when the loop is most visible in this channel. In Figure 7 we display the light curves in all 6 channels over 1.2 hours at a representative pixel on the loop (located near the arrowhead in Figure 6). The light curves in each channel are all normalized to their own maxima, and each curve is offset by 0.5 in y relative to the previous one. In order from bottom to top, we plot light curves from channels 131, 171, 193, 211, 335 and end with 94. Clear dynamic behavior is evident in the light curves of all of the channels; there is an obvious rise and fall in the intensity on a time scale of ~ 20 minutes (though in 94 this behavior is comparable to variations due to noise). This dynamic behavior where the intensity rises and then falls is not consistent with steady heating. Moreover, the peak intensity in the different channels is reached in sequentially cooler channels as time progresses, qualitatively consistent with the predictions from the nanoflare storm models discussed earlier. The 335 light curve peaks at ~8.2 UT, 211 and 193 both peak around ~ 8.35 UT, and 171 and 131 both peak around ~ 8.4 UT.

Next, we use a rigorous background subtraction method to isolate the loop (loop_width3.pro, available in SolarSoft; Klimchuk 2000). In this method the loop is identified by the user, straightened, and the background is determined by linear interpolation across the axis of the loop, at each point along the loop axis. We compute the light curves by integrating along the length (11 pixels) and width (4 pixels) of the background subtracted loop and display them in Figure 8. Additionally, we have done an 11-pt temporal smoothing (~11 minutes) to highlight the overall rise-and-fall behavior of the loop and minimize the higher frequency contributions.

The general behavior of the single-pixel, no-background-subtracted case (a ~ 20 minute rise and fall in the intensity) is exhibited in this rigorous, loop integrated, background subtracted case. The 335 light curve still reaches its peak at ~ 8.2 UT, followed by 211, 193, 094, 171 and then 131. In these light curves we see that, though close to each other, 211 clearly precedes 193 by a few minutes throughout the duration of the loop evolution. Additionally, the signal to noise of channel 94 improves significantly. Now it is evident that the 94 emission rises and falls somewhat, peaking around 8.1 UT, and then peaks again with larger emission at 8.4 UT. Referring back to the lower intensity EBTEL example shown in Figure 3a, we see that the small rise in 94 is predicted to occur first as a result of hot emission, followed sequentially by peaks in the light curves of 335, 211, 193, then the main peak of 94, due to the cooler emission, and finally, 171 and 131. The light curves of this loop are consistent with heating by a nanoflare storm of lower intensity and short duration, and are not consistent with steady heating. We point out that the EBTEL example used here as a comparison was a standard run (Example 1 from Klimchuk et al. 2008), and we made no attempt to match the exact properties. Another run using the same nanoflare energy and a more appropriate loop half-length of $3.5 \times 10^9$ cm (based on the images) gives much shorter time delays (e.g., 0.3 hr between 335 and 171), in better accord with the observations. Furthermore, the current version of EBTEL tends to underestimate the rate of cooling in the late cooling phase, which could also account for some of the differences between the details of the observed light curves and the predicted light curves. We are developing an improved version, which should be available shortly (Cargill, Bradshaw, & Klimchuk, in preparation).

We note that the apparent thickness difference of the loop in the 171 and 335 images of Figure 6 is an illusion. Because the loop is a relatively smaller intensity enhancement relative to the diffuse emission in 335, it appears to be thicker. However, rigorous examination of the background subtracted loop shows that the width is similar in 335 and 171. There are documented cases where more than one warm loop occupies the envelope defined by a thicker hot loop, though not necessarily at exactly the same time (Winebarger & Warren 2005). Guarrasi et al. (2010) have offered a possible explanation which reconciles all of these observations.

3.2 Location 2: Loop Emission

For the second location, we again choose a loop which is visible in both 171 and 335 in the active region core, however the behavior of the light curves of this loop are, at first glance, less consistent with nanoflare storm heating than the loop at location 1. We display linearly scaled images of the core in 171 and 335 (Figure 9), magnified 5 times, and arrows to identify the loop of interest. The 171 image was taken at 14.6 UT, while the 335 image was taken a little after 14 UT. In Figure 10, we plot 4-hour normalized light curves of a representative pixel in the loop; as in Figure 7, we begin with 131 on the bottom, followed by 171, 193, 211, 335 and 94, each offset in y by 0.5. There is an overall rise and fall in the 335 emission, with the peak intensity occurring a little after 14 UT. The light curve of channel 171 peaks around 14.6 UT, after 335 as we would expect given the nanoflare storm models presented in Figure 3; however, the 171 light curve also has a significant peak around 13.6 UT, well before the peak in 335. Examination of 211 and 193 also reveal this two-peaked behavior, with significant intensity peaks both before and after the 335 intensity peak. This event is clearly not described by steady heating; however, it exhibits features which are inconsistent with the simple nanoflare storm heating scenarios we showed in Figure 3.

Particularly in an active region core, lines of sight are likely to intersect many strands and even many loops. In general, these loops or strands are behaving independently, and with different properties such as total energy release. This can result in more complicated light curves than we observed at location 1. Given probable line of sight overlap, we might reasonably expect that this loop in location 2 has significant contributions from two or more independent loops, especially since we made no attempt at a background subtraction. Furthermore, certain line of sight geometries will preclude the researcher from ever fully isolating emission due to an individual loop in an image; therefore, it is valuable to understand what we would expect to see if two physically separate nanoflare storms occurred in separate loops along the same line of sight.

To this end, we revisit EBTEL for the case of two independent nanoflare storms along a line of sight. In this case, we initiate a storm of weak nanoflares, followed after a 4000s delay by a storm of nanoflares which have ten times as much energy per strand (nanoflare examples 1 and 4 of Klimchuk et al. 2008). Both storms last 3500s, and there are 3 times as many strands activated in the storm of weak nanoflares. After convolving the resulting emission with the AIA response functions, we get the result shown in Figure 11, where each light curve is normalized to its own maximum. The peak order in this case is a combination of the different magnitude storms and the AIA response functions. Just as we observe at location 2, the 211, 193 and 171 channels all have peaks before the 335 channel, even though 335 is a hotter channel. Upon closer inspection, the model 335 channel light curve does have two peaks, one associated with each nanoflare storm. The first one however is small; given the lower signal to noise ratio of the 335 channel, it would be easy to miss in the actual data. This first 335 peak does occur before the peak in 211 does, just as we expect for cooling nanoflare heated strands. Since the second storm is stronger than the first, the 335 peak associated with the second storm is much brighter than the peak associated with the first. This explains why we do not see a clear peak in 335 preceding the first 211 peak.

The observed light curves of 211, 193 and 171 all have their second peaks occurring after the 335 peak, in that order. Additionally, though the 094 channel is noisy, it exhibits a clear peak around 13.9 UT, just after the first 171 peak, and just before the 335 peak, consistent with the model predictions. Comparing the light curves observed at location 2 with the model results, we conclude that this event is consistent with two nanoflare storms occurring on two physically separate loops successively along the line of sight, with the first one being weak, and the second one being strong. The agreement is rather impressive, since, as before, we made no attempt to fine tune the model.

3.3 Location 3: Diffuse Emission

Lastly, we choose a pixel in the active region core at a location of diffuse emission, a location where there appears to be no discrete loops. Recall from our discussion of Figure 4 that the diffuse emission between loops in the AR core comprises the majority of the emission of the AR in all of the channels. As we discuss in the Introduction, it is important to understand if the heating which produces this 'diffuse' background emission is fundamentally different than the heating of loops. Is this diffuse emission best described by steady heating, or is it also explained by impulsive nanoflare heating? If it is described by nanoflare heating, is it a storm of nanoflares where some mechanism creates physical coherence, organizing nearby strands to initiate nearby in time, or are the nanoflares occurring on random, unrelated strands? We look for a region where there is no discernable loop in any of the 6 channels, and identify such a region at 10 UT. In Figure 12, we plot the active region core, magnified 5 times, displayed on a linear scale, in all 6 channels at 10 UT. The box indicates the diffuse region that we chose based on the six images at this time.

We plot 3-hour light curves at a pixel located in this diffuse emission in the center of the box outlined in Figure 13. As with the loop locations, these light curves exhibit dynamic behavior, namely the rise and fall of intensity in all of the channels; in this instance the enhancement is longer, ~ 90 minutes, rather than 20-30 minutes as in the loops examined. As we discussed for locations 1 and 2, this dynamic behavior is not consistent with steady heating, or emission from plasma which is in hydrostatic equilibrium. The peak enhancement in the light curves are not reached at the same time, rather the peaks of this long scale variability are reached at 335 at 9.7 UT; 094 around 9.9 UT; 211 around 10 UT; 193 10.2 UT; and 171 and 131 around 10.3 UT. We also plot the 30-minute smoothed light curves for channels 335, 211, 193 and 171, normalized to their own maxima, in Figure 14 to illustrate this long timescale dynamic behavior more clearly. In the smoothed light curves it is clear that the intensity rises and falls with a time lag between channels, once again consistent with the temporal behavior of a nanoflare storm. The duration of the enhancement in the light curves indicates that the storm lasts for well over an hour (Klimchuk 2009) and the delays between the channels of the 90-minute envelope enhancement suggests that the strands are long. At the time of the images shown in Figure 12 (10 UT), the emission due to the long duration nanoflare storm is fading in 335, peaking in 211 and 193, and rising in 171. Though the behavior is consistent with nanoflare storm heating, as in the loop examples, the total emission at 10 UT is not significantly brighter than neighboring pixels in any channel, so we identify it as background or diffuse emission. The light curves for this location in all of the channels suggest that this long-duration (~90 minutes) nanoflare storm comprises a significant portion of the diffuse emission. As we discussed in Section 3.3, some of the total emission observed in this diffuse, intra-loop region could be due to scattering from nearby bright features such as other loops or moss (e.g. DeForest et al. 2009). Importantly, we have examined all nearby bright features in these images, and none of them exhibit temporal variability correlated with the temporal variability we focus on here. We can be confident that the temporal variability arises from emitting plasma located along the line of sight of this location.

For all of these locations we analyze, we demonstrate that the temporal behavior is consistent with nanoflare storm heating. It is important to know whether the count rates observed in the active region core are also consistent with nanoflare storm heating. Bradshaw and Klimchuk (2011) predict count rates in several AIA channels for the loop apex of fourteen different nanoflare models computed with the 1D HYDRAD code (Bradshaw & Cargill 2010 and references therein), taking full account of nonequilibrium ionization. The 195/335 ratio ranges between 13 and 446 and has a median value of 46. In comparison, the average 195 and 335 count rates in a 30x20 pixel area in the core of this observed active region have a ratio of 11. Bradshaw and Klimchuk (2011) did not determine count rates for the 171 channel. The overall magnitudes of the intensities observed in this active region core are consistent with the models. Detailed comparisons will be addressed a later paper.

## 4. Summary and Conclusions

We present an investigation of the heating of NOAA Active Region 11082 using the wonderful new observations made with the SDO/AIA telescopes. We demonstrate that in this active region there is plenty of warm (~1 MK) emission in the 'hot' core of the active region, even when taking into account possible foreground emission. In fact, in the (warm) 171 channel emission in the core is brighter than emission in the more traditionally thought of warm 'extended loop' region. Though the intensity slice of the hot emission falls off more quickly outside the core region than the warm emission does, hot and warm emission are not mutually exclusive. Also, while discrete loops are features which subjectively stand out in images, loops are at most ~ 35% of the total emission along a given ling of sight in any of the channels, and often are only ~10%. Diffuse emission dominates in the entire active region.

We examine light curves of three locations in the active region core, two loop regions and one diffuse region, and demonstrate that the behavior observed at all three locations can be described in terms of nanoflare storm heating. None of the three location's light curves are consistent with steady heating. This result is consistent with earlier studies of warm, 'extended' loops surrounding active region cores (e.g. Ugarte-Urra, Winebarger, Warren 2006; Ugarte Urra, Warren and Brooks, 2009; Winebarger and Warren, 2005; Winebarger Warren and Seaton 2003). While nanoflare storm heating of warm extended loops has been previously modeled and its time-lag signatures observed, we expand here to demonstrate that the light curves in the *core* of this active region are also consistent with nanoflare storms. Clearly loops occur both in active region cores as well as in the volume surrounding cores, and they are described by a common mechanism (nanoflare storms). In light of these facts, we suggest abandoning the term 'extended loops', and calling any discrete enhancement a loop, regardless of whether it occurs in the core or in the 'extended' region surrounding the core.

Furthermore, we feel that the result for the diffuse location is especially significant. We presented an instance where diffuse emission is consistent with a long duration storm of nanoflares. This suggests that the physically-organizing mechanism which concentrates nanoflares into nearby strands and generates visually identifiable loops may also be generating some portion of diffuse emission in active regions. Is all diffuse emission caused by storms of nanoflares which happen not to produce a distinct loop, as we saw in our example? Alternatively, is there a component of the diffuse emission which is heated by nanoflares which are randomly initiated with no organized, physical coherence? Research is currently underway to address this question.

Additionally, we demonstrate that seemingly out of order light curves are also consistent with nanoflare storm heating. The second location we present is better understood as a combination of two nanoflare storms, a weak one followed by a strong one, which we model with EBTEL. Given line of sight effects, this scenario is expected to occur often in active region cores. In Figure 15 we show a summary of the order in which the AIA light curves are predicted to peak for the three nanoflare storm scenarios discussed in this paper. Though there are different peak orderings with different storm parameters, the possible light curves are understandable and predictable. Note that the 211-193-171 order is maintained for all three of these models. The light curves of the three locations taken together demonstrate a possible common physical mechanism, impulsive nanoflare heating, operating in at least some coronal loops and some regions of the diffuse corona. We speculate that nanoflares are ubiquitous. This possibility is supported by recent studies of the distributions of soft X-ray intensity fluctuations (Terzo et al. 2011) and of widespread plasma of very high temperature (Reale et al. 2009). Whether the light curves presented here are an anomaly or represent typical conditions can only be answered with a rigorous statistical analysis of the behavior of the entire active region; such a study is currently underway.

Acknowledgements

The research of NMV was supported by an appointment to the NASA Postdoctoral Program at the Goddard Space Flight Center, administered by Oak Ridge Associated Universities through a contract with NASA. The research of JAK was supported by the NASA Supporting Research and Technology program. The data are courtesy of NASA/SDO and the AIA science team.

Figure Captions

**Figure 1** Active region observed on 2010 June 19, at 8.4 UT using SDO/AIA. All 6 images are coaligned, and displayed on a linear scale. Left to right, top to bottom, channels 131, 171, 193, 211, 335 and 94 Å. Temperature of peak sensitivity are listed in the top left corner. White lines indicate slice used for Figure 4; 1 pixel = 0.6".

**Figure 2** Normalized SDO/AIA temperature response functions for EUV channels 131 (black), 171 (cyan), 193 (orange), 211 (blue), 335 (green) and 94 (red). Same color scheme used in all figures.

**Figure 3** AIA intensities predicted with EBTEL due to a 500 second nanoflare storm. a) low intensity nanoflare storm. b) a high intensity nanoflare storm.

**Figure 4** Intensity profile along center of active region in 171 and 335. Diagonal slice indicated with white lines in Figure 1. 171 emission measure is multiplied by 52.

**Figure 5** 171 image (as shown in Figure 1) with foreground regions indicated with white rectangles.

**Figure 6** Location1. Magnified core region in 171 (image taken at 8.4 UT) and 335 (image taken at 8.2 UT).

**Figure 7** Location 1. Normalized light curves of a representative pixel on the loop. From bottom to top are plotted 131, 171, 193, 211, 335, and 94, each offset by 0.5 in y.

**Figure 8** Location 1. Background subtracted, loop-integrated, normalized light curves, smoothed with an 11-minute running average.

**Figure 9** Images of location 2. Magnified core region in 171 (image taken at 14.6 UT) and 335 (image taken at 14 UT).

**Figure 10** Location 2. Normalized light curves of a representative pixel on the loop, as displayed in Figure 7 for location 1.

**Figure 11** AIA intensities predicted due to EBTEL-modeled 3500 second weak nanoflare storm followed by a 3500 second strong nanoflare storm.

**Figure 12** Images of location 3. Magnified core region at 10 UT, for all 6 EUV channels used. Box indicates location 3.

**Figure 13** Normalized light curves at representative pixel in location 3, as displayed in Figure 7 for location 1.

**Figure 14** Normalized, 30-minute running average of light curves observed in 335, 211, 193 and 171 at location 3. Long-scale variation exhibited in all channels, with time lags between them.

**Figure 15** Summary figure of light curve peak orders for different nanoflare storm scenarios.

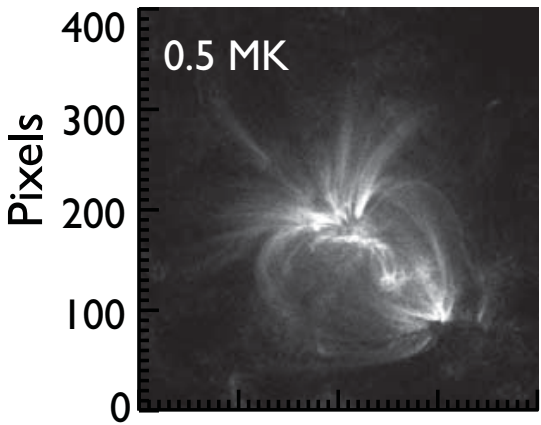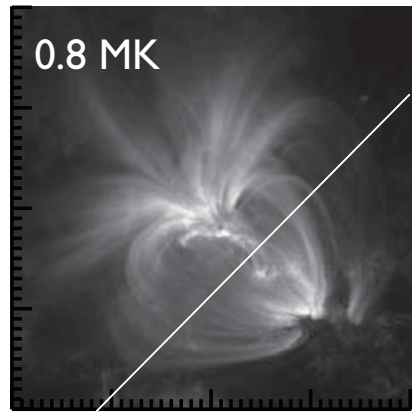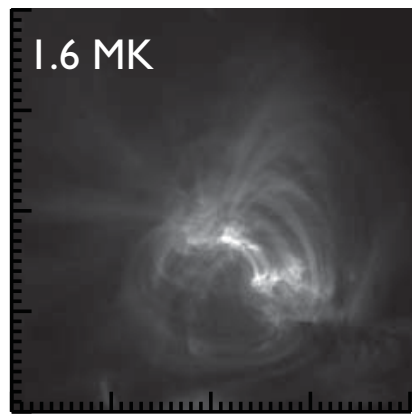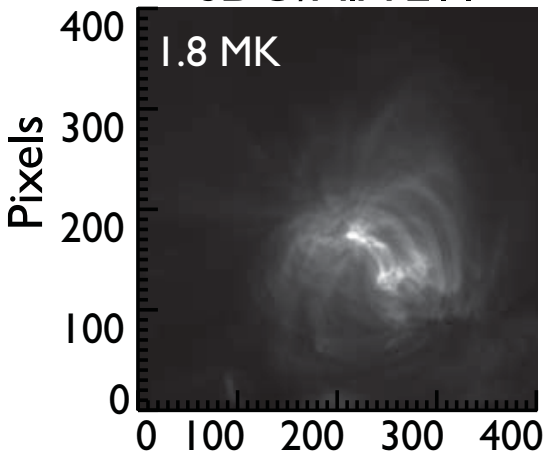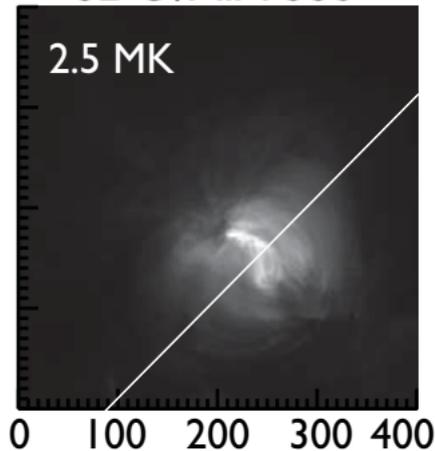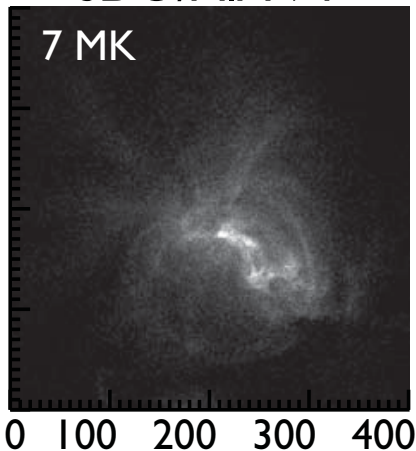

Figure 2

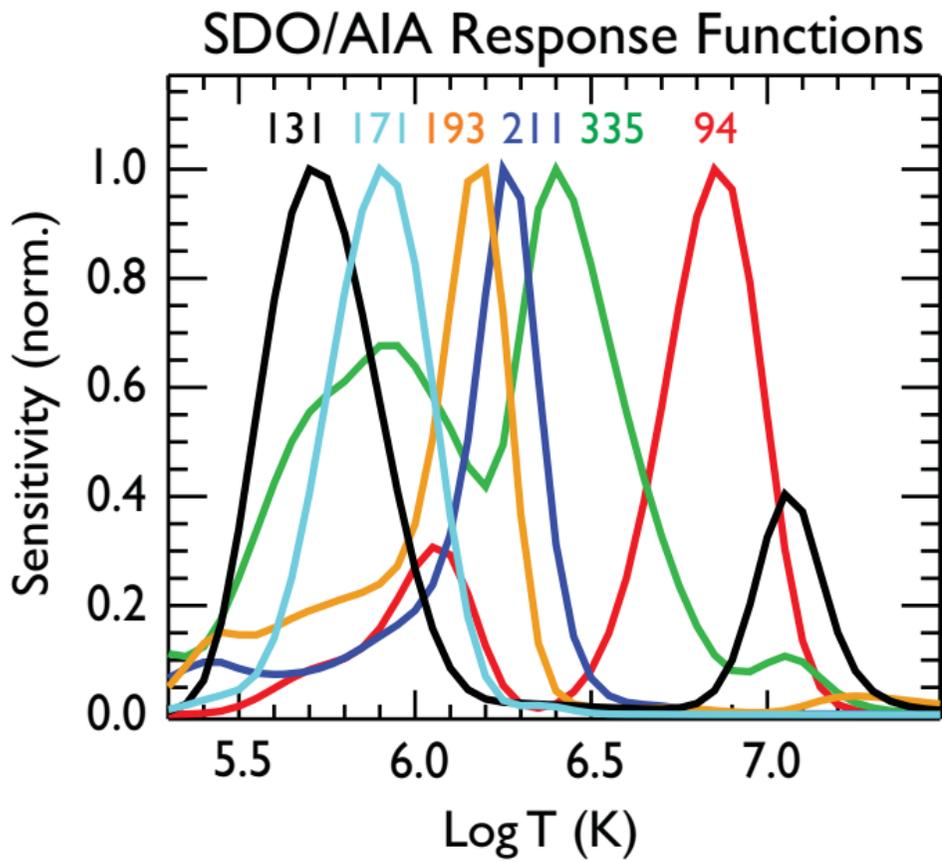

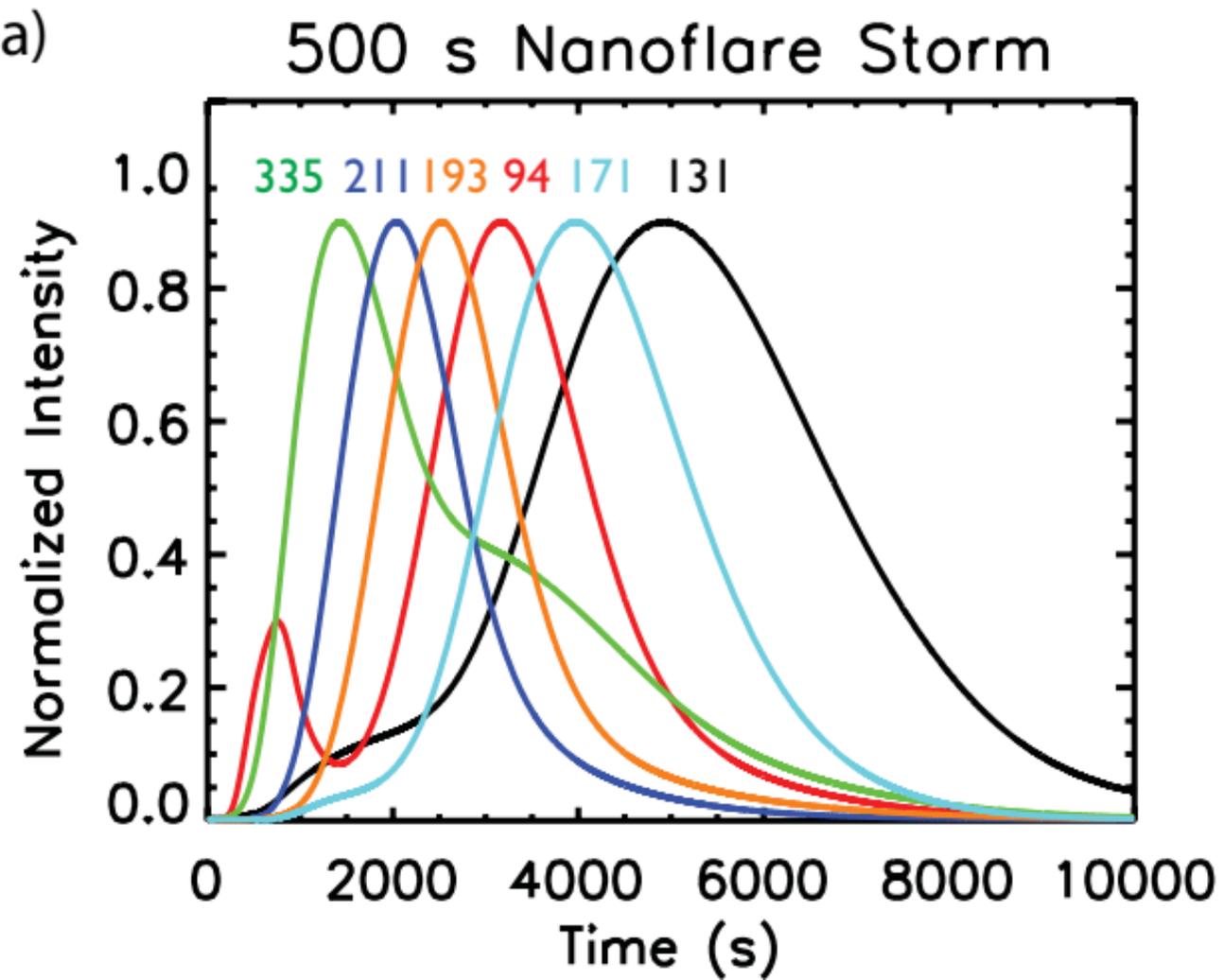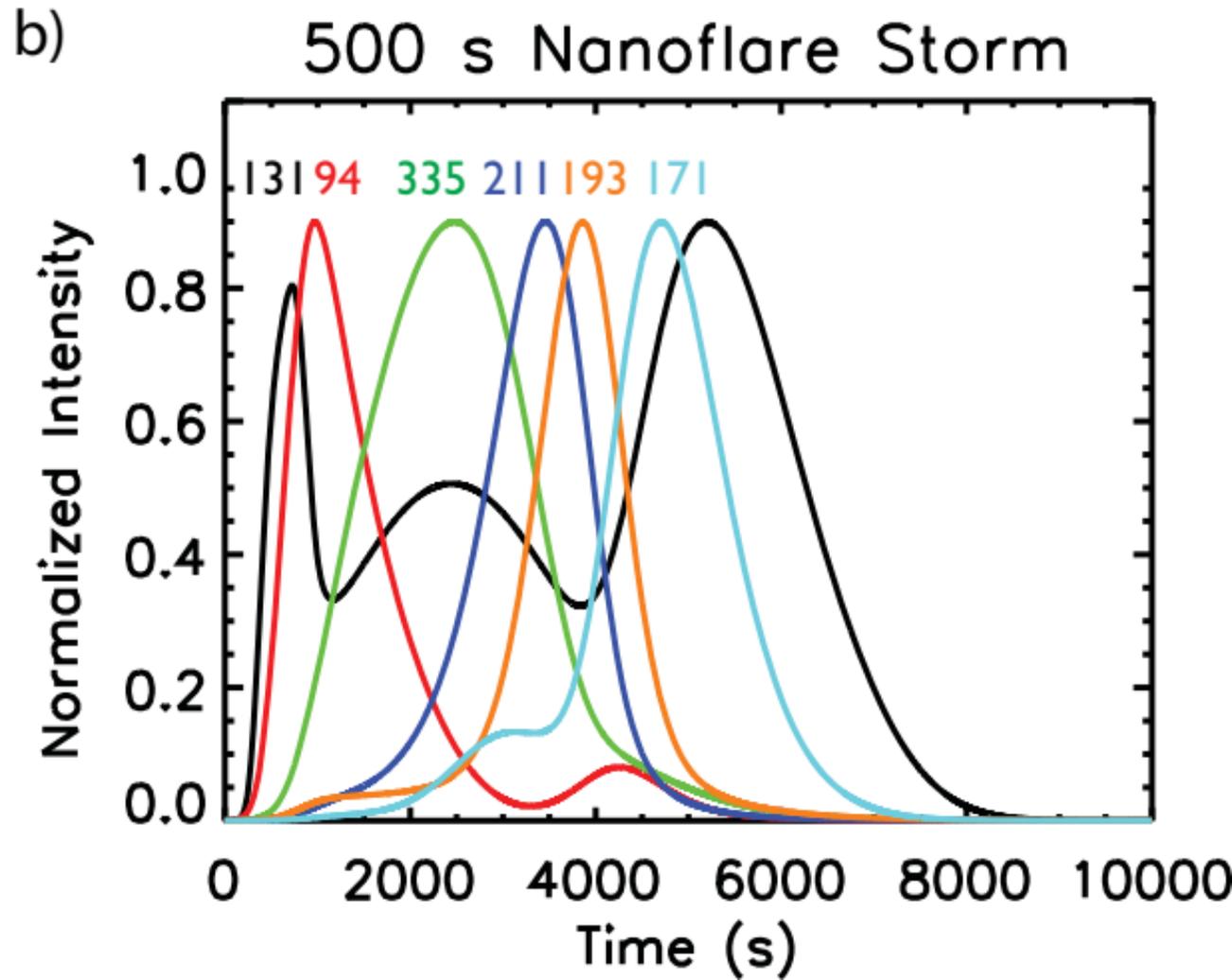

Figure 4

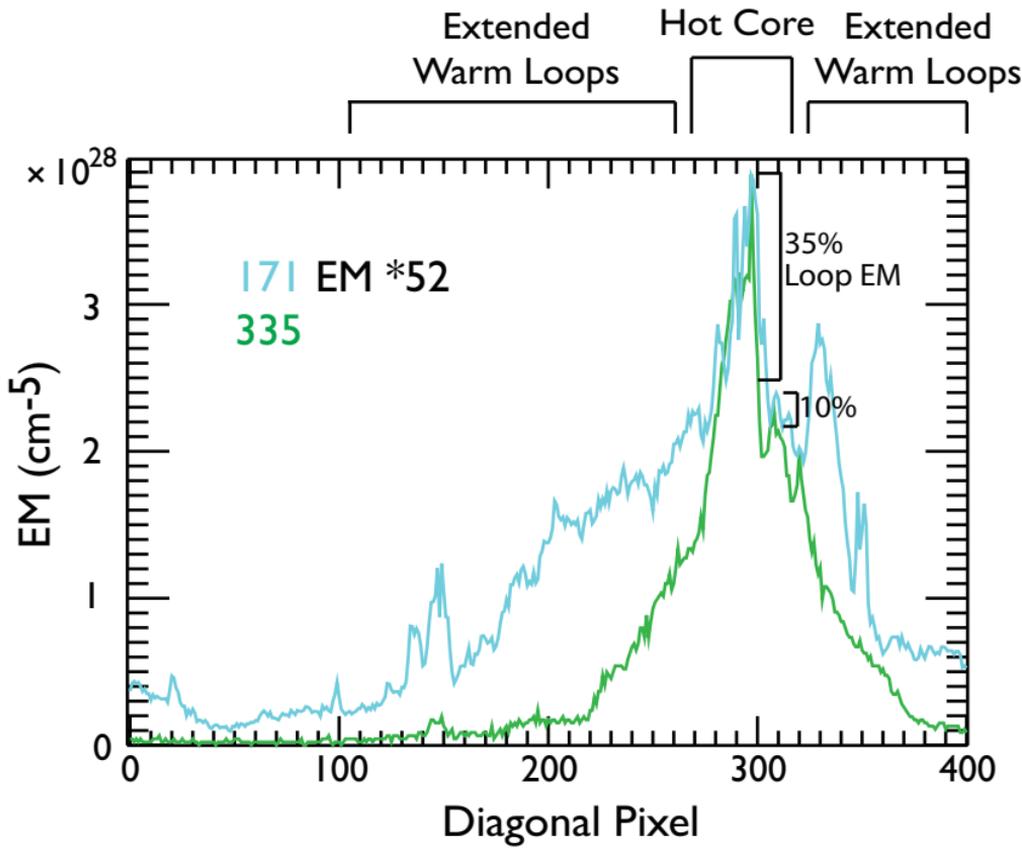

Figure 5

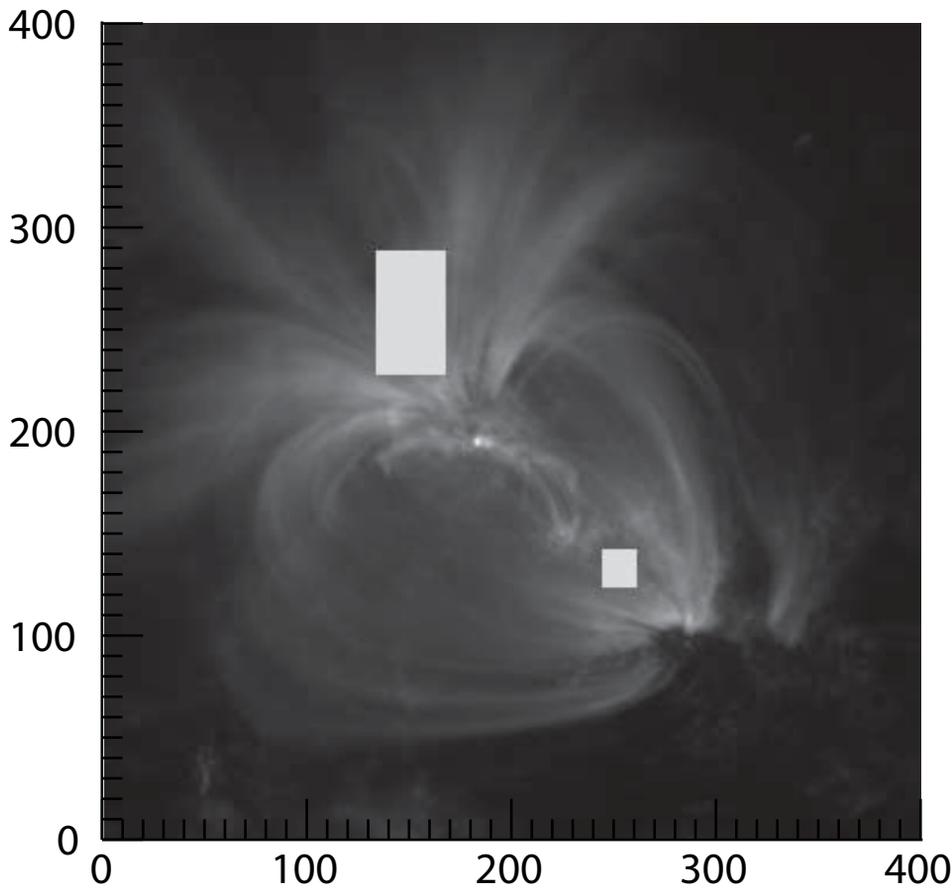

Figure 6

## Location 1, AR Core

### 171

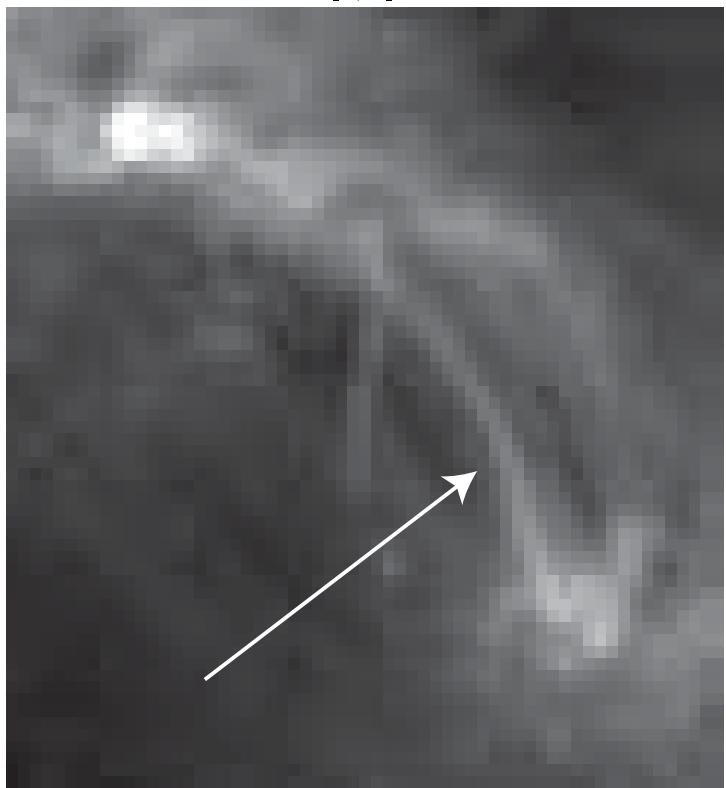

### 335

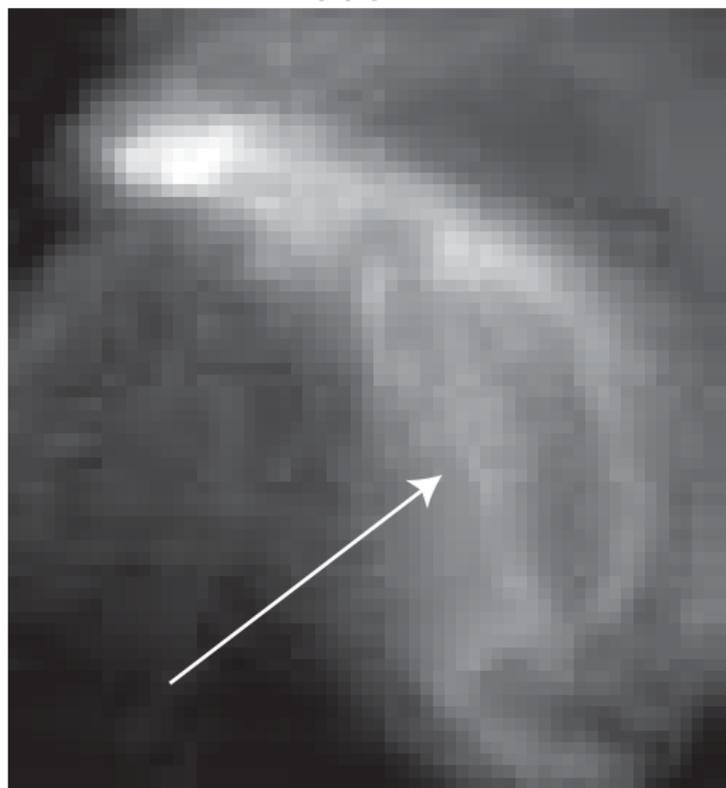

Figure 7

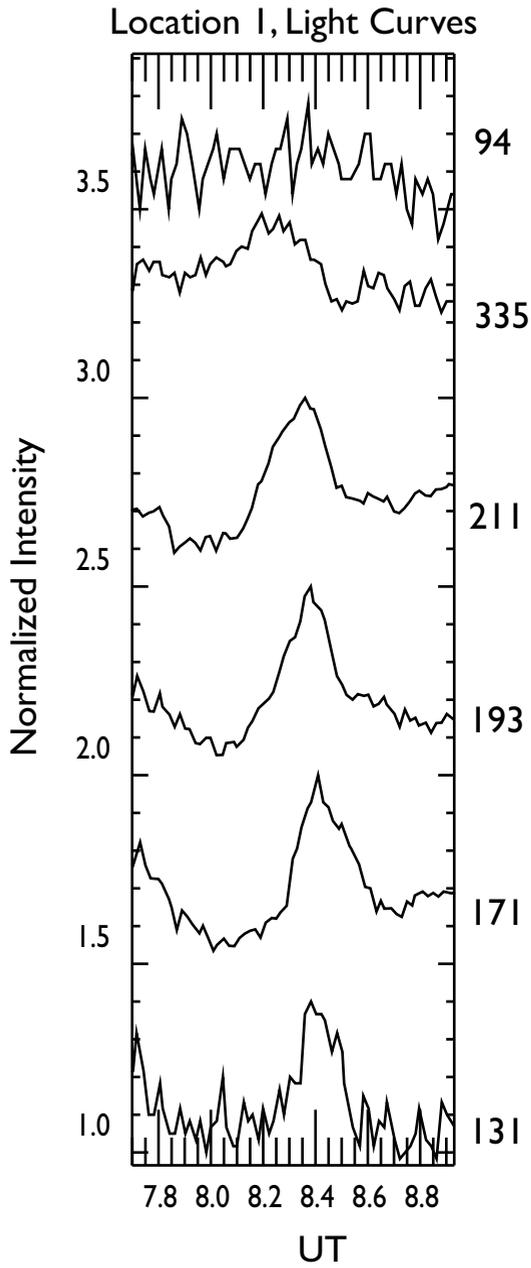

Figure 8

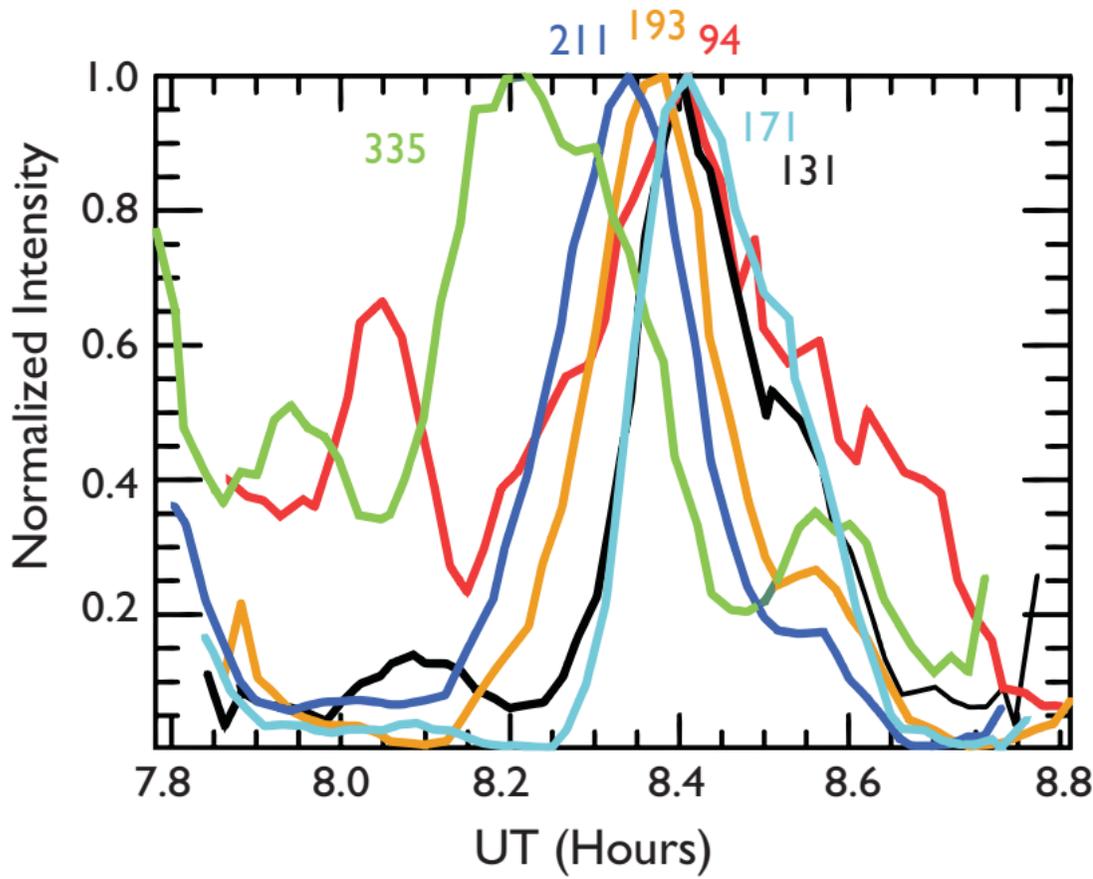

Figure 9

## Location 2, AR core

### 171

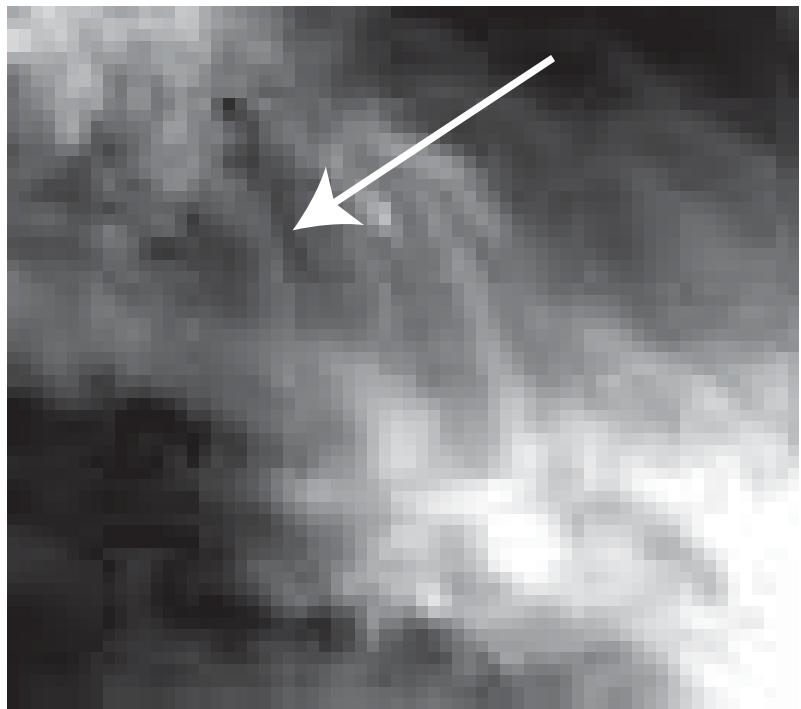

### 335

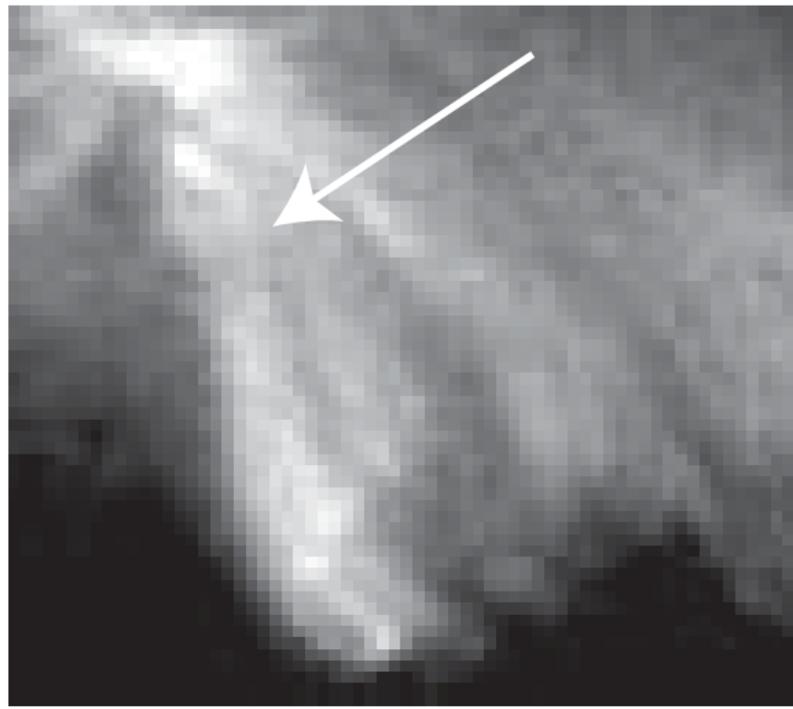

Figure 10

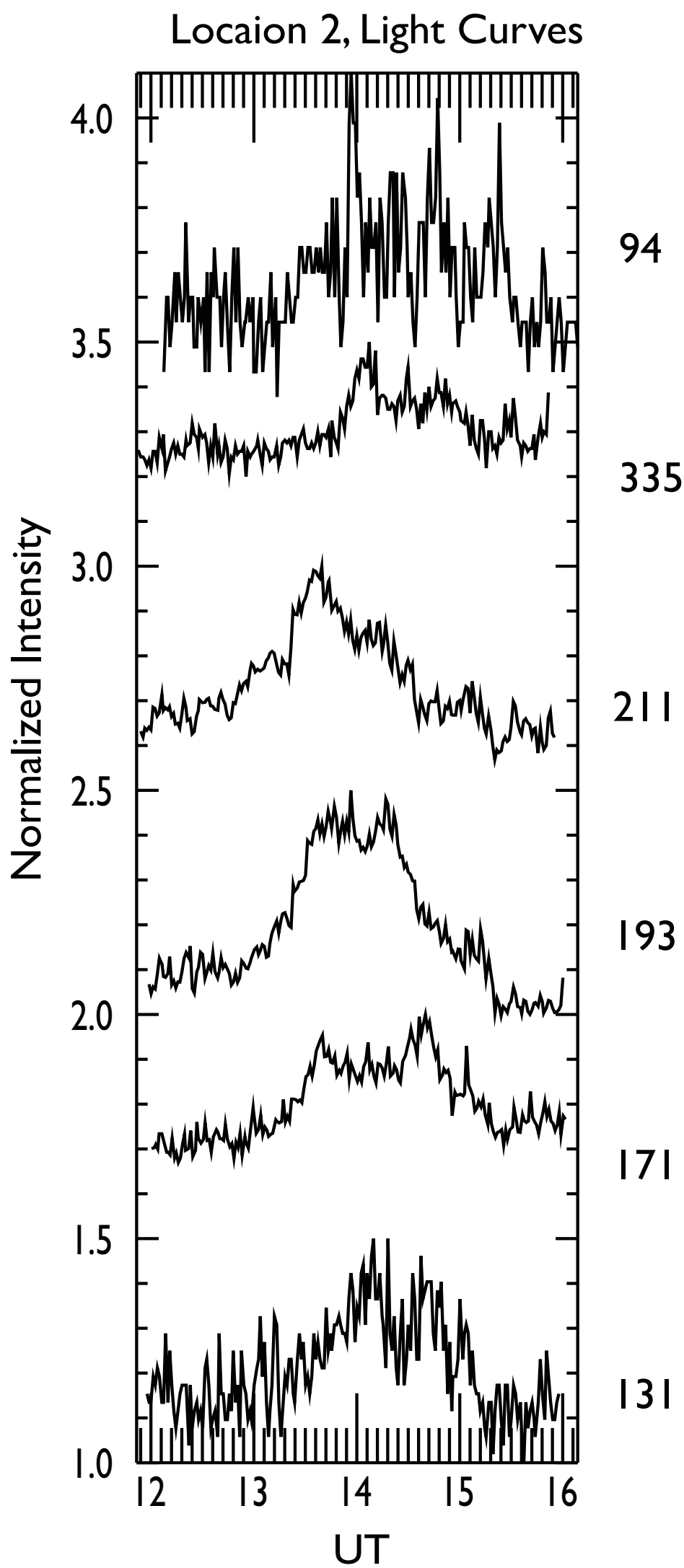

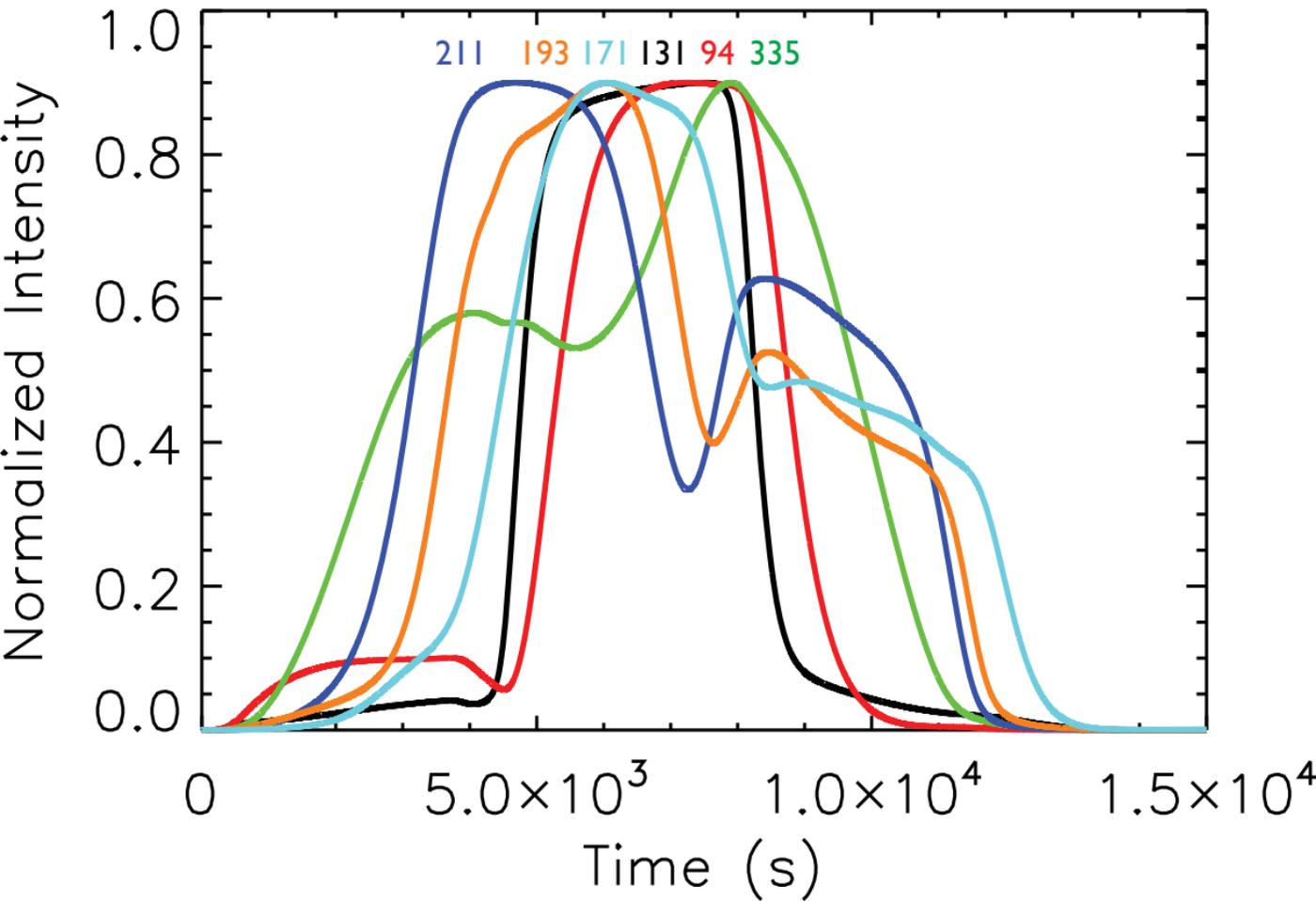

Figure 12

# Location 3, AR Core Images

### 131
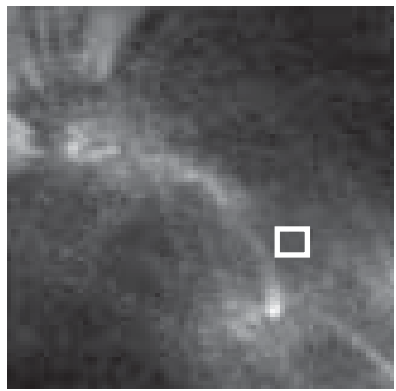

### 171
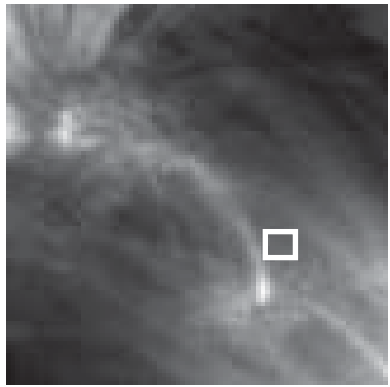

### 193
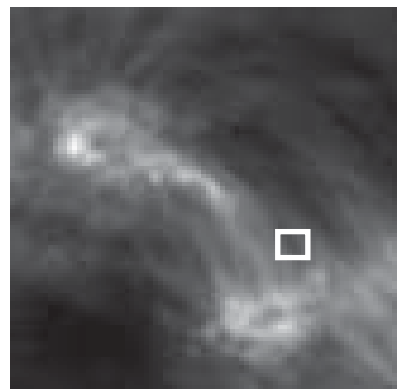

### 211
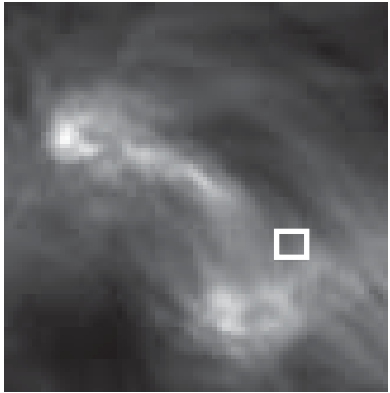

### 335
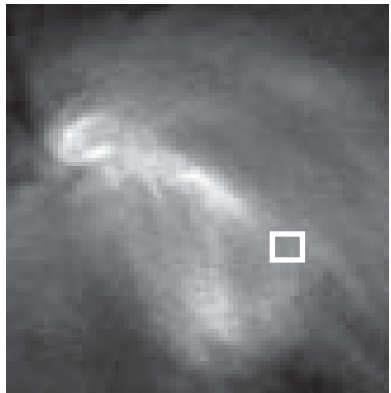

### 94
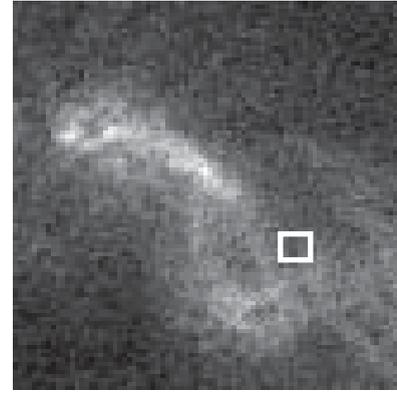

Figure 13

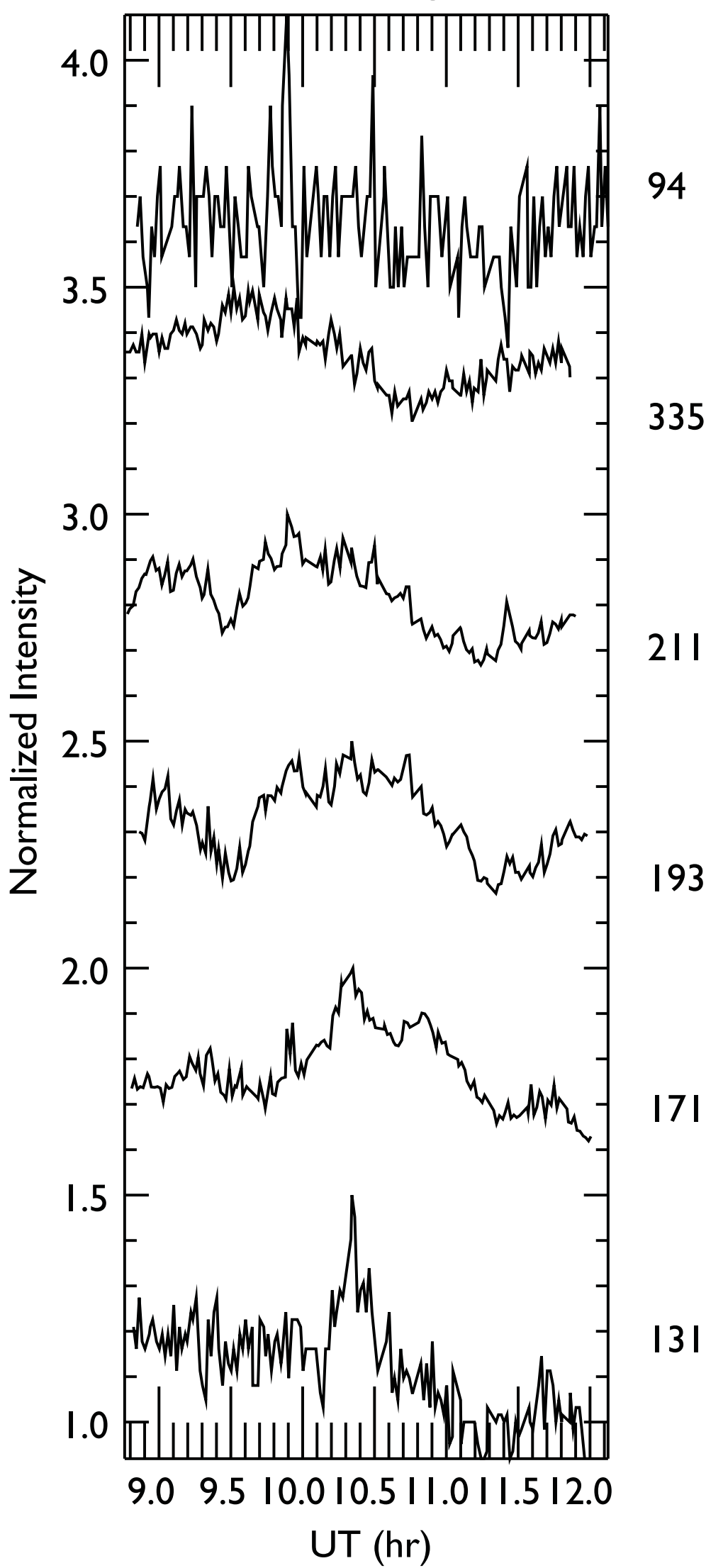

Figure 14

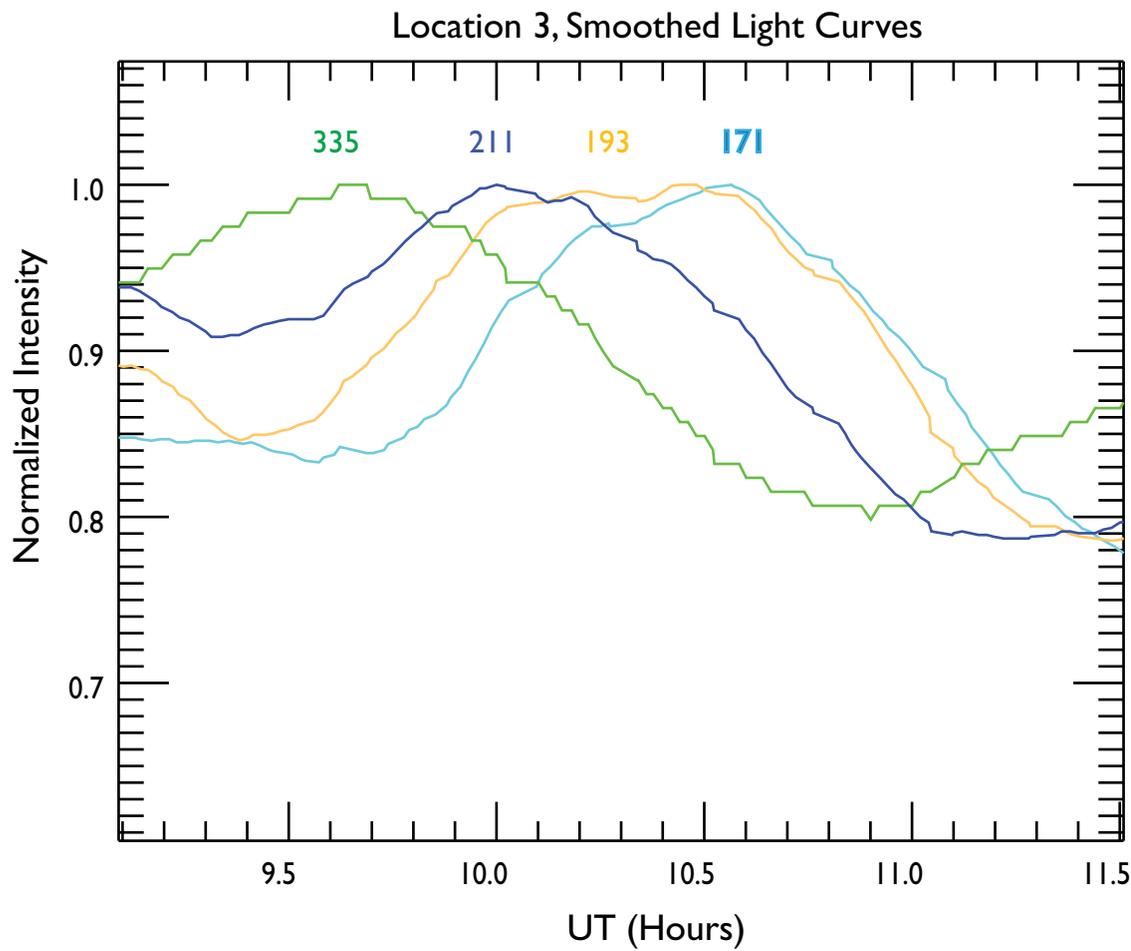

## Order of Peak Intensities

| | | | | | | |
|---|---|---|---|---|---|---|
| Strong, 500s | 131 | 94 | 335 | 211 | 193 | 171 |
| Weak, 500s | 335 | 211 | 193 | 94 | 171 | 131 |
| Weak then Strong, 3500s | 211 | 193 | 171 | 94 | 131 | 335 |